# Implementation of EMR System in Indonesian Health Facilities: Benefits and Constraints

1st Muhammad Rasyid Juliansyah
PT Medigo Teknologi Kesehatan
Jakarta, Indonesia
rasyid@klinikpintar.id

2nd Bukhori Muhammad Aqid
University of Indonesia
Jakarta, Indonesia
bukhori.muhamad@cs.ui.ac.id

3th Andien Putri Salsabila
PT Medigo Teknologi Kesehatan
Jakarta, Indonesia
andien@klinikpintar.id

4th Kurnia Nurfiyanti
PT Medigo Teknologi Kesehatan
Jakarta, Indonesia
nia@klinikpintar.co.id

*Abstract*—This white paper delves into the widespread implementation of Electronic Medical Records (EMR) within healthcare facilities across Indonesia. It examines the driving forces behind EMR adoption, particularly the role of government regulations, and addresses the challenges encountered by clinic owners and healthcare providers in transitioning to these digital systems. Furthermore, this paper highlights the significant benefits and transformative advantages of EMR systems, such as enhanced decision-making through real-time data access (around 15-20 minutes time saved for patient waiting time and approximately saved 20-25 minutes for all service duration), reduction in healthcare costs over time due to improved resource management, and increased patient satisfaction by providing faster and more personalized care. EMR systems also ensure higher levels of data security and privacy, adhering to national healthcare standards, while supporting continuous monitoring and updates that enhance system resilience and functionality. The findings are substantiated through case studies, such as case study at LAPAS II Purwokerto Clinic and case study at PMI Purbalingga Clinic and user testimonials from clinics that have successfully implemented EMR solutions in compliance with the standards established by the Ministry of Communication and Informatics (Kominfo) and the Ministry of Health (Kemenkes).

## I. INTRODUCTION

The healthcare sector in Indonesia is undergoing significant transformation with the adoption of Electronic Medical Records (EMR). This shift is largely driven by government regulations aimed at improving healthcare delivery, enhancing data accuracy, and ensuring patient safety. This white paper examines the chronology of EMR implementation, the challenges encountered by clinics and the overall benefits derived from EMR adoption.

## II. BACKGROUND

The implementation of EMR in Indonesian healthcare facilities is mandated by the Ministry of Health, following regulations aimed at modernizing healthcare practices. The Ministry of Communication and Informatics (Kominfo) has set standards for electronic system providers, ensuring that EMR systems are secure, reliable, and efficient.

### A. Key Regulations:

- Ministry of Health Regulation Permenkes 24 about Medical Records.
- Ministry of Communication and Information Regulation No. 20 of 2016 on Personal Data Protection in Electronic Systems.
- Ministry of Communication and Information Regulation No. 5 of 2020 on Electronic System Organization (PSe)

These regulations emphasize the importance of transitioning from paper-based records to digital systems to enhance healthcare quality and operational efficiency.

## III. CHALLENGES FACED BY CLINIC OWNERS AND HEALTHCARE PROVIDERS

### A. Complexity and Cost

The complexity of system integration arises from the need to ensure seamless communication between the EMR and existing systems. This involves not only technical compatibility but also the harmonization of data formats, protocols, and interfaces [1]. For instance, disparate systems may use different data formats, requiring extensive data mapping and transformation to ensure that information flows correctly between systems. Additionally, ensuring that these systems can securely exchange patient data in real-time, while maintaining data integrity and compliance with regulatory standards, adds another layer of complexity.

The cost implications are significant. Integration projects often require specialized IT expertise, which can be expensive. Moreover, the process may involve upgrading or replacing outdated systems, further increasing costs [2]. According to a study by the Healthcare Information and Management Systems Society (HIMSS), healthcare providers often underestimate the financial and time resources needed for successful EMR integration. The costs can quickly escalate due to unforeseen challenges such as software incompatibilities or the need for additional custom development.

*B. Data Migration*

Data migration is another significant technical challenge in EMR implementation. It involves transferring patient records from paper-based systems to digital formats. This process is not only time-consuming but also requires meticulous attention to detail to ensure that all patient data is accurately captured and transferred without loss or errors [3].

TABLE I. CATEGORIES AND DURATION OF DATA MIGRATION

| Type | Amount Data | Amount Clinic | Duration |
|---|---|---|---|
| Patient | 285,178 | 48 | Oneday |
| Service | 3693 | 40 | |
| Inventory | 4024 | 45 | |

Internal data of Klinik Pintar

*C. Maintenance Costs*

Even after an EMR system is implemented, clinics must deal with ongoing maintenance costs, which can also strain their budgets. These costs include regular system updates, technical support, data storage, and cybersecurity measures to protect sensitive patient information [2] . Over time, the cumulative expense of maintaining an EMR system can be significant, particularly for clinics that have limited financial resources.

*D. Technical Support and Training*

To ensure that the EMR system operates smoothly, clinics often need ongoing technical support from the software vendor or a third-party IT service provider. This support can include troubleshooting issues, performing routine maintenance, and providing additional training as new features are introduced [2]. The cost of these services is typically billed on a subscription basis, adding a recurring expense to the clinic's budget.

Additionally, as staff turnover occurs or as the system evolves, continuous training is necessary to keep healthcare providers up to date with the latest functionalities of the EMR system. These training sessions, while crucial for maintaining system efficiency, also contribute to the overall maintenance costs [4].

*E. Comprehensive Training Requirements*

EMR systems are complex, involving multiple functionalities such as patient data entry, medical history tracking, billing, and reporting. To ensure healthcare providers can navigate these systems confidently, training programs need to cover a broad range of topics, including system navigation, data entry protocols, patient confidentiality practices, and troubleshooting common issues [5].

Training programs must be tailored to the different roles within a healthcare facility. For instance, doctors, nurses, administrative staff, and billing personnel may all interact with the EMR system differently and therefore require role-specific training. This segmented approach ensures that each group can fully understand and utilize the aspects of the EMR system that are most relevant to their duties.

*F. EMR Adoption Impact on Smaller Clinics*

For smaller clinics, these initial costs can be particularly prohibitive. Unlike larger hospitals or healthcare networks, small clinics typically operate with tighter margins and have less access to capital. The financial strain of implementing an EMR system may require these clinics to seek external financing or reallocate resources from other critical areas, such as patient care or facility maintenance [6].

In some cases, smaller clinics may delay the implementation of EMR systems due to cost concerns, opting instead to continue using traditional paper-based records. However, this decision can put them at a disadvantage in terms of efficiency, data management, and regulatory compliance, particularly as digital health records become increasingly standardized across the healthcare industry.

According to internal data Aplikasi Klinik Pintar, health facilities categorized as Independent Doctor's Practices that operate independently, without any collaboration with third parties, predominantly utilize Electronic Medical Record (EMR) tools that follow a freemium-based model. This choice is largely attributed to the user-friendly nature of these tools, which comply with the standards set by PERMENKES 24.

Interestingly, only about 2% of Doctor's Practices take advantage of patient Data Migration services. This limited usage can be attributed to the substantial amount of information contained in physical medical records, making the transition to digital format a time-consuming process. As such, many practices find it challenging to undertake this migration efficiently.

*G. Data Storage and Cybersecurity*

The digital nature of EMR systems necessitates secure data storage solutions that comply with national regulations on patient data protection. This often involves investing in cloud storage services or on-premise data centers, both of which come with associated costs [7]. Furthermore, clinics must implement robust cybersecurity measures to safeguard against data breaches, which can be costly to set up and maintain. These measures may include firewalls, encryption, multi-factor authentication, and regular security audits.

IV. CASE STUDIES AND USER TESTIMONIALS

*A. Case Study 1: Implementation EMR at LAPAS II Purwokerto Clinic*
- *Timeline*
    - On March 28 2023, Klinik Pratama Lapas Kelas II A Purwokerto officially received a clinic operational permit
    - The processes of documenting, reporting, managing, and storing patient information at Klinik Pratama Lapas Kelas II A Purwokerto are conducted manually.
    - During the period of 2022-2023, the Ministry of Health of Indonesia implemented compulsory regulations

- pertaining to the utilization of the Electronic Medical Record (EMR) system.
- In September 2023, a socialization event regarding Electronic Medical Records (EMR) was conducted, facilitated by the Ministry of Communication and the Ministry of Health. This event featured an EMR provider, specifically Klinik Pintar.
- As a result of the socialization of Electronic Medical Records (EMR) conducted by Klinik Pintar, Klinik Pratama Lapas Kelas II A Purwokerto expressed interest in the implementation of EMR utilizing the Klinik Pintar application.

- Challenges
    - Regulatory Compliance
    As regulations mandate the implementation of Electronic Medical Records (EMR), healthcare facilities must transition from traditional manual record-keeping to electronic systems promptly and effectively.
    - Complexity and Cost
    The expenses associated with developing a custom application, along with the necessity for ongoing server maintenance and system enhancements, present significant obstacles for medical clinics.
    - Ensuring Data Quality and Integrity
    Clinics face challenges associated with traditional record keeping and data management, which are labor-intensive and susceptible to inaccuracies.

- Results
    - Regulatory Compliance
    Clinics are not required to allocate substantial funds for the development and maintenance of specialized applications; by utilizing the RME provided by AKP, they gain access to a reliable and consistently updated electronic medical record (EMR) platform.
    - Improved Efficiency
    Following comprehensive training sessions on the use of AKP, conducted both online and offline, which were held multiple times to guarantee that all personnel at the clinic adeptly acclimate to the new system and implement it effectively within their respective departments.

    *"Since implementing Klinik Pintar, monitoring the health of our inmates has become significantly more efficient. I can easily access medical records, track disease progression, and generate accurate reports for evaluating rehabilitation programs. This has greatly assisted me in fulfilling my duties as the head of the inmates' guidance section."* –Mr. Fauzen, Head of Inmate Guidance Section at Purwokerto Class IIA Penitentiary Clinic

    - Streamlined Workflow
    The introduction of the EMR system markedly decreases the manual workload, ensuring that all data is organized and accurately documented. During a recent drug audit at the clinic, the findings received commendations for the management of the prison clinic, attributed to the effective implementation of the RME system, which facilitated meticulous record-keeping.

TABLE III. COMPARISON OF BEFORE AND AFTER IMPLEMENTING THE EMR AT KLINIK PRATAMA LAPAS KELAS IIA PURWOKERTO

| Change | Before June 2023 | July - December 2023 |
|---|---|---|
| System Usage | Manual System | Electronic System (Klinik Pintar App) |
| Patient Waiting Time | 20 - 30 minutes from registration made | 5 - 10 minutes from registration entered through the system |
| Patient Service Duration | 40 - 50 minutes from admission to discharge | 15 - 20 minutes from admission to discharge |
| Expense Allocation Paper & Office Supply | ~3 million IDR/month (printing medical records and prescription copies) | No longer printing medical records and prescription copies |

Internal data of Klinik Pintar

B. *Case Study 2: Implementation EMR at PMI Purbalingga Clinic*
- Timeline
    - Klinik PMI Purbalingga was established in 1999, initially known as the PMI Treatment Center. In 2013, in accordance with government regulations, the clinic officially changed its name to Klinik PMI Purbalingga.
    - Discussions about the transition to electronic medical records (EMR) began in 2020, sparked by a quarterly meeting of the Indonesian Doctors Association (IDI) that assessed the benefits and necessities of RME.
    - Although EMR is not yet mandatory, regulations from the government are starting to encourage the adoption of this

system in healthcare facilities, such as Klinik PMI Purbalingga.

- Challenges
  - Ensuring Data Quality and Integrity
    Before switching to the EMR system, Klinik PMI Purbalingga depended on handwritten medical records. This approach faced challenges in accessing data and created a threat of losing medical information.
  - Resistance to Change
    Certain healthcare professionals showed reluctance to adopt the EMR system, favoring the traditional methods they are accustomed to.
  - Perceived Workflow Disruptions
    Healthcare professionals were worried that utilizing the EMR system could impede workflow, particularly in the initial phases of its rollout.
- Results
  - Regulatory Compliance
    Through the implementation of the EMR system, PMI Clinic achieved adherence to governmental regulations while simultaneously improving the protection and privacy of patient information with a more cohesive and secure framework.
  - Improved Efficiency
    After the introduction of the new system, PMI Clinic saw a notable improvement in its operational effectiveness. The processes for patient registration and record management were streamlined, leading to shorter wait times (reduced around 15-20 minutes per patient) and fewer mistakes in handling patient information.
  - Quick Access to Patient Records
    The EMR system allows healthcare personnel to swiftly and effortlessly retrieve patient medical records without having to look for physical files. This not only expedited the provision of services but also improved the precision of the information being documented.

*"Being part of a healthcare organization that has implemented Electronic Medical Records (EMR), I have witnessed considerable advantages. The EMR system enhances service speed, minimizes documentation errors, and decreases paper consumption along with storage requirements. These enhancements contribute to increased efficiency in work processes and greater comfort for patients at PMI Clinic in Purbalingga Regency."* – Dr. Retno, Director of PMI Clinic in Purbalingga Regency.

V. BENEFITS AND ADVANTAGES OF EMR IMPLEMENTATION

The transition to EMR systems offers numerous benefits for clinics and healthcare providers:

*A. Improved Efficiency*

Implementing Electronic Medical Records (EMR) systems brings significant improvements in the efficiency of healthcare operations. By transitioning from paper-based records to digital systems, healthcare providers can streamline their workflows, reduce administrative burdens, and focus more on patient care [2].

TABLE IV. DURATION OF ACTIVITY IN HEALTHCARE

| Activity | Duration | |
|---|---|---|
| | Before EMR | After EMR |
| Registration | More than 5 mins | Less than 3 mins |
| Wait for Consultation* | More than 10 mins | 2-3 mins |
| Consultation** | More than 15 mins | Average in 10 mins |
| Wait for Payment*** | More than 10 mins | Less than 3 mins |

Internal data of Klinik Pintar

*) after the previous patient leaves the consultation room
**) if common disease consultation
***) if patient is not prescribed from doctor

EMRs dramatically reduce the time healthcare providers spend on administrative tasks, such as filing, searching for, and updating paper records. Instead of dealing with cumbersome paperwork, providers can quickly input, retrieve, and update patient information electronically. This not only speeds up administrative processes but also minimizes the risk of lost or misplaced records. With automated reminders, scheduling systems, and streamlined documentation, EMRs enable healthcare providers to manage their time more effectively, allowing them to dedicate more attention to direct patient care [8].

*B. Quick Access to Patient Records*

One of the most immediate advantages of EMR systems is the ability to access patient records instantly. In a traditional paper-based system, retrieving a patient's file can be time-consuming, especially if the records are stored off-site or are not well-organized. EMRs allow healthcare providers to pull up complete patient histories with just a few clicks, which is particularly beneficial in emergency situations where time is of the essence. Quick access to accurate patient information can significantly improve response times in medical emergencies, leading to better patient outcomes [5].

*"In the past, patient identification was done based on their village or neighborhood unit, not by alphabet or number as it is now. This is somewhat different from the modern system which is more universal and structured. However, even though the system is simple, this method is quite effective in identifying which village patients come from. Since switching to electronic medical records with*

*Klinik Pintar, access to medical data has become much faster and more efficient. This application is integrated with Satu Sehat and BPJS, making it easier for me to obtain patient medical records without any obstacles. Using Klinik Pintar is also very helpful in fulfilling regulations, accreditation processes, and ensuring that practices run smoothly. This allows me to focus more on health services without being distracted by manual data management."* – Dr. Yaziruddin

*C. Error Reduction*

Manual data entry, often required in paper-based systems, is prone to human error, such as misinterpretation of handwriting, incorrect data entry, or incomplete records. EMRs mitigate these risks by providing standardized forms, automated data validation, and error-checking mechanisms. These features help to minimize the incidence of data entry errors, ensuring that healthcare providers have access to accurate and reliable patient information at all times [1] [7]. This reduction in errors is particularly important in contexts where even small mistakes can have serious consequences for patient health.

The transition to EMR systems also enhances data accuracy, which is critical for providing high-quality healthcare. By digitizing patient records and standardizing data entry processes, EMRs reduce the risks associated with manual data handling and ensure that patient information is consistently accurate and up-to-date.

According to the journal 'Computerized physician order entry can decrease medication errors by more than one half, although not all of these errors would have resulted in an adverse event. In other research, a computerized physician order entry system helped reduce non missed-dose medication errors in outpatients from 142 per 1000 patient-days at baseline to 26.6 per 1000 patient-days after the intervention'[9].

*D. Consistency*

EMR systems enforce standardized data entry formats, which contribute to greater consistency across patient records. Consistent data is easier to analyze and interpret, allowing healthcare providers to make more informed decisions based on complete and uniform information. Standardization also facilitates better communication between different departments and care providers, as everyone is working with the same data formats and terminologies. This consistency is essential for maintaining high standards of care and ensuring that all healthcare providers are on the same page when it comes to patient information [10].

*E. Better Patient Care*

EMRs significantly enhance the quality of patient care by providing healthcare providers with comprehensive and easily accessible patient histories [8]. This comprehensive view of patient information supports more informed medical decision-making and improves coordination among healthcare providers.

*"Our patients are very satisfied with the service because they no longer need to bring a paper prescription. They just wait for their medication, making the process much simpler and more convenient for them."*-Klinik Mawar

*F. Comprehensive History*

One of the key benefits of EMRs is their ability to compile a complete and accessible patient history. All relevant medical information, including past diagnoses, treatments, medications, allergies, and lab results, is stored in one place. This comprehensive view allows healthcare providers to make more informed decisions about treatment plans and to identify potential issues that might not be apparent from isolated pieces of information [11]. For example, a provider can quickly see how a patient has responded to previous treatments, helping to tailor future care to the patient's specific needs.

*G. Data Security*

In an increasingly digital world, the security of patient data is paramount. EMR systems are built with robust security features, including encryption, access controls, and audit trails, to protect sensitive patient information from unauthorized access and data breaches [12]. Compliance with Kominfo's regulations on data security ensures that healthcare facilities are safeguarding patient data in line with national standards, thereby reducing the risk of costly breaches and protecting patient trust.

VI. CONCLUSION

The implementation of Electronic Medical Records in Indonesia is a critical step towards modernizing the healthcare system. Despite the challenges, the benefits in terms of efficiency, data accuracy, reduced medication error and improved patient care make EMR adoption a worthwhile investment. By learning from international examples and tailoring the approach to local needs, Indonesia can achieve significant advancements in healthcare delivery.